\documentclass[preprint,3p,times,twocolumn]{elsarticle}

\usepackage{amssymb}

\usepackage{dcolumn}
\usepackage{bm}
\usepackage{amsmath}
\usepackage[T1]{fontenc} 
\usepackage{xcolor}
\usepackage[version=3]{mhchem} 
\usepackage{booktabs}

\journal{Molecular Liquids}

\begin{document}

\begin{frontmatter}



\title{Water in an electric field does not dance alone:
 The relation between equilibrium structure, time dependent viscosity and  molecular motions}


\author[PULS,EAM]{Andreas Baer}
\author[PULS,EAM,IRB]{Zoran Mili\v{c}evi\'{c}}
\author[EAM,IRB]{David M. Smith}
\ead{dsmith@irb.hr}
\author[PULS,EAM,IRB]{Ana-Sun\v{c}ana Smith\corref{COR}}
\ead{smith@physik.fau.de}

\cortext[COR]{Corresponding author}

\address[PULS]{PULS Group, Institute for Theoretical Physics, FAU Erlangen-N\"urnberg, 91058 Erlangen, Germany}
\address[EAM]{EAM Cluster of Excellence, FAU Erlangen-N\"urnberg, 91052 Erlangen, Germany}
\address[IRB]{Division of Physical Chemistry, Ru\dj er Bo\v skovi\'c Institute, 10000 Zagreb, Croatia}

\begin{abstract}
Dynamic structuring of water is a key player in a large class of processes underlying biochemical and technological developments today, the latter often involving electric fields. However, the anisotropic coupling between the water structure and the field has not been understood on a molecular level so far.  Here we perform extensive molecular dynamics simulations to explore the influence of an externally imposed electric field on liquid water under ambient conditions. Using self-developed analysis tools and rigorous statistical analysis, we unambiguously show that water hydration shells break into subcompartments, which were hitherto not observed due to radial averaging. The shape of subcompartments is sensitive to the field magnitude and affects excitations of the hydrogen bond network including the femtosecond stretching and the sub-picosecond restructuring of hydrogen bonds. Furthermore, by analysing the reorientational dynamics of water molecules, we ascertain the existence of cooperative excitations of small water clusters. Enabled by the interplay between hydrogen bonding, and the coupling of water dipoles to the field, these coordinated motions, occurring on the picosecond time scale, are associated with fluctuations between torque-free states of water dipoles. We show that unlike the coupling between translation and reorientation of water molecules, which takes place on even longer time scales, these coordinated motions are the key for understanding the emergent anisotropy of diffusion and viscosity of water. Particular effort is invested to provide an analysis that allows for future experimental validation

\end{abstract}


\end{frontmatter}


\section{\label{sec:Intro}Introduction}
Despite extensive research efforts, liquid water continues to surprise the scientific community with its unusual qualities, both as a bulk material and as a solvent \citep{Meng}. Interestingly, a number of its anomalous properties are of structural origin \cite{Nilsson}. Namely, unlike in simple liquids, where interactions between constituents are isotropic, the water molecule is polar, which among other things makes it capable of directly interacting with local electric fields \cite{Evans1987}. Furthermomre, each water molecule has the capacity to participate in up to four, partially covalent, hydrogen (H)-bonds, which spontaneously stabilize transient and localized molecular arrangements \cite{Errington}. 

The dynamics of H-bonding actually plays a crucial role for the structure of water. As revealed by a number of experimental \citep{Hura20001,Hura20002,Soper1997,Soper2000} and simulation \citep{Lee2006,Todorova,Kuhne2009} studies, the coordination of first neighbours around the central water molecule yields nearly four H-bonds being formed on average. Besides the well established symmetric tetrahedral structure, disordered motives have been recently identified to contribute to the structure, in a temperature independent manner \cite{Gallo2016,Morawietz2018}. The dynamics of the network is governed by the interplay between diffusion and kinetics for the formation of H-bonds, with a life time of about a picosecond \citep{LuzarNat}. This interplay gives rise to correlation functions with a non-exponential decay with a characteristic time of about a few tenths of a picosecond \citep{Luzar2000}, which is a time scale systematically appearing in scattering experiments \citep{Teixeira,Laenen,Nienhuys} and in molecular dynamics (MD) simulations \citep{LuzarNat,Luzar2000}.

High resolution X-ray spectroscopy data, however, has implicated the existence of a significantly faster relaxation process related to perturbations of the order and symmetry of the H-bond network \citep{Wernet2004,Tokushima2008,Tokushima2010,Tokushima2012}. While highly debated over the last decade, the emergent picture is that, on femtosecond time scales, a single water molecule participates in one strong donor and acceptor bond, while the secondary donor and acceptor bonds are significantly weaker. The origin of this asymmetry was recently explained by fast thermal distortions of H-bonds, most likely being at interplay with stochastic rotations of the molecules occurring on these time scales \citep{KuhneNat,KuhneJACS}.

Both of these processes (slow and fast) contribute to thermally excited stress relaxations within the fluid, and have a clear signature in both pressure-pressure correlation functions and the associated viscosities \cite{Guo,Medina2011,Phillips,Li,Laslett,Staib,Shekhar,Milischuk}. These correlation functions are concerned with the evolution of ensemble characteristics. Besides being experimentally available, these global characteristics converge with computationally reasonable efforts. The downside of such calculations is, that associating a particular relaxation with molecular degrees of freedom may be a challenging task. Nonetheless, the time evolution of viscosity is often used to extract particular characteristics of typical relaxations in fluids, including their functional type and relaxation time scales \cite{Heyes,Vallauri,Marti,Guo}. Hence, the analysis of stress relaxation may be a valuable tool for the detection of new relaxation processes in systems where these may occur. 

One such system is water under the influence of electric ($E$)-fields. Namely, a uniaxial electric field destroys the symmetry of the  Hamiltonian, thus making the appearance of new relaxations in the laboratory frame of reference possible \cite{Evans1987}. Actually, it was argued already more than 30 years ago that time cross-correlations between molecular linear and angular velocity can acquire non-zero values under the influence of the field \cite{Evans1984}. Since the field-induced torques act anisotropically and primarily on rotational degrees of freedom, alterations of diffusion coefficients, which are non-homogeneous \cite{Evans1987,Vegiri2004}, must be associated with rotational-translational coupling \cite{Evans1986}. This finding was corroborated somewhat later with the observation that slow translational motions coincide with slow completion of tumbling cycles of individual molecules in supercooled liquid water \cite{Vegiri2004}. This effect should depend strongly on the strength of the field and could be detectable in stress relaxations. Nonetheless, despite the importance of this coupling for the development of new technologies \cite{Shui,Brutin,Choi,Gencoglu,Mugele}, no systematic analysis of the relaxation processes in water, under the influence of $E$-fields in ambient conditions, has been performed so far. 

Until now, most work addressed changes in the two named relaxation processes both in clusters  \cite{Vegiri2001a,Vegiri2001b, Vegiri2002, Vegiri2003, Chakraborty2018}  and bulk water \cite{Vegiri2004, Hu2011} supercooled to 70-240 K. Here, even relatively weak $E$-fields ($E<0.5$ V/nm) induced a strong anisotropy in the reorientational correlation function and the intermediate scattering function. An increase of long-time relaxations along the field axis was observed, although the average relaxation times remained virtually unaffected in the bulk systems. The excess of field-induced torque was associated with enhanced reorientational motion promoted by an increased frequency of restructuring of H-bonds and an increase of the vibrational amplitude within the cage, resulting in a blue shift of the  O-O-O bending vibration of around 50 cm$^{-1}$ \cite{Vegiri2004}. Increasing the field strength to 1.2 V/nm was found to lock the water dipoles along the field axis, allowing only for spinning around the field, which is however, restricted by immediate neighbours. Hence, the decays of correlation functions became strongly stretched \cite{Vegiri2004}. This finding is consistent with recent ab-initio MD (AIMD) simulations \cite{Futera2017}, even though there are controversial discussions about whether such effects can be caught by classical MD simulations \cite{English2015}.

At field values above 1 V/nm water was found to undergo a structural transition to an amorphous state, where the tetrahedral structure is strongly compromised. Nonetheless, the onset of the locking effect was reported already at 0.6 V/nm, where a dramatic increase of self-diffusion coefficients and relaxation rates of translational and rotational degrees of freedom were reported \cite{Vegiri2001a}. These changes were associated with a strengthening of H-bonds along the field direction, which seemed to promote freezing \cite{Vegiri2004}. However, due to the strong restriction imposed by the low temperature, it is unclear if such behaviour persists under ambient conditions and for larger systems, which experience smaller pressure fluctuations, especially since the described effects seem to be more pronounced at lower temperatures.

At room temperature, it was suggested that the freezing transition follows a path similar to that in supercooled water \cite{Sutmann} but the density correlation functions, calculated in the frame of the water molecule, did not show any significant modification with the field \cite{Zong2016}. On the other hand, the anisotropy of self-diffusion coefficients \cite{Evans1987, Sutmann, Kiselev1996}, related to the viscosity at $t\rightarrow\infty$ \cite{Zong2016} has been well established over the past decades. Consistently with a more ordered medium \cite{Vegiri2004, Jung1999, Zong2016}, the self-diffusion parallel to the field axis was found to decrease upon increasing the field strength. Along the perpendicular direction it was shown for the SPC/E water model, that an initial increase is followed by a decrease in the self-diffusion coefficient, with a maximum at around 0.6 V/nm \cite{Zong2016}. However, it has yet to be determined if this long-time limit behaviour of self diffusion and viscosity is dependent on the choice of the water model. An even more pertinent open question concerns the nature and the characteristic time scales of the relaxation processes, under the influence of the $E$-field, at room temperature. 

It is precisely these issues that we address in the present contribution. Building on our previous experience with simulating complex liquids \cite{Milicevic,Brkljaca}, we perform extensive molecular dynamics simulations of water in $E$-fields (for methods see section 2). Using  unprecedented statistical sampling and rigorous methodology, we characterize the effect of the field on the static density correlation functions (section 3) and the time evolution of the shear viscosity for moderate fields (sections 4 and 5). By coupling this analysis with an extensive investigation of the orientational dynamics, we are able to unambiguously connect distinctive molecular excitations with macroscopic stress relaxations (section 6). This allows us to identify a new, cooperative process  on the picosecond time scale, which clearly couples the reorientation  of water molecules belonging to the same first hydration shell. The experimental consequences of our results are discussed in the concluding section (7) of the manuscript.

\section{\label{sec:Methods}Computational Methods}
\textbf{\emph{Simulation setup:}}
We performed extensive MD simulations of bulk water, in the absence and presence of an external electric field, within the GROMACS simulation package (versions 4.5.5 and 5.1.4) \cite{GROMACS, Berendsen1995, Lindahl2001, Hess2008, Goga2012, Pronk2013, Abraham2015}. The $E$-field was applied along the laboratory $x$-axis using the tinfoil boundary condition along the field axis \cite{Hunenberger}. All simulations were performed with a time step of 1 fs and periodic boundaries along axes where there is no field. We used the LINCS \cite{LINCS} algorithm for restraining the length of the covalent OH bond and the HOH angle. Van der Waals interactions were treated by the shifted 12-6 Lennard-Jones potential \cite{Christen} (shift from 0.9 to 1.2 nm), whereas for electrostatic interactions the smooth particle-mesh Ewald technique was used at separations larger then 1.2 nm \cite{PME, Essmann1995}.

\textbf{\emph{Equilibration protocol:}}
For every setup, the equilibration of the system was achieved through  $\approx7$ ns simulations in the NPT ensemble, with target pressure and temperature of  $P =1$ bar and $T=300$ K, respectively. The desired temperature was maintained with the Nos\'e-Hoover thermostat \cite{Nose,Hoover}, whereas the pressure was controlled using the Parrinello-Raman barostat \cite{Parrinello1,Parrinello2}. NPT runs were sufficiently long to allow for electrostriction (equilibration of density and energy), the result of which is the linear increase of water density with the $E$-field shown in the Supplementary Information (SI) SI-Fig.~1a.  Equilibration was completed by 10 ns  simulations in NVT ensemble which were started using a configuration for which the pressure was 1 bar and the volume equivalent to the average volume from the last nanosecond of the NPT run. All production runs were performed in the NVT ensemble.

\textbf{\emph{Water models:}}
To allow for comparison with results in the literature \cite{Evans1987, Evans1984, Vegiri2004, Evans1986, Vegiri2001a, Vegiri2001b, Vegiri2002, Vegiri2003, Sutmann, Zong2016, Kiselev1996,Jung1999} three commonly used rigid water models were explored, namely TIP4P \cite{TIP4Porig}, SPC/E \cite{SPCE}, and TIP4P/2005 \cite{TIP4P2005}. At this stage, flexible force fields were not considered since they are associated with considerable computational cost and because for  fields below 3 V/nm the changes in the dipole moment of water are less than $5\%$ \cite{Sutmann}. Furthermore, since the applied electric field is 1 to 10\% of intrinsic electric fields in water, which are between 15 and 25 V/nm \cite{Jung1999,Sutmann,English2003}, polarizable water models, which are even more computationally demanding, were not explored.

\textbf{\emph{Static structure:}}The static structure was explored in simulations with $N_{\text{w}}=1000$ water molecules in $E$-fields in the range of $0-3.2$ V/nm. All simulations were performed following the above described protocol with a total length of the production run of 10 ns. The analysis was performed with a self-developed software for calculating a two-dimensional density distribution function that accounts for the azimuthal symmetry introduced by the field.

\textbf{\emph{Transport coefficients:}}
Anisotropic diffusion coefficients were determined from the trajectories generated to study the static structure. The analysis was performed with a self-developed routine (using \cite{Tange2018}) on an ensemble of eight times 1000 trajectories with a length of 2.5 ns each. Fitting to the linear regime of the time evolution of the mean square displacement, averaged over all trajectories in one set, provided a distribution of transport coefficients. The average of this distribution is associated with the diffusion coefficient under fixed conditions, while the variance is taken as the uncertainty. This method yields significant error bars challenging the statistical accuracy of small changes in diffusion constants.      

In complex fluids, the diffusion can be expected to deviate from pure Gaussian. To resolve these deviations, we calculate the time evolution of the non-Gaussian parameter $\alpha_{\textbf{x}_i} (\Delta t)$
\begin{equation}\label{eq:nongauss}
\alpha_{\textbf{x}_i} (\Delta t) = \frac{1}{5}\frac{\langle \Delta \textbf{x}_i (\Delta t)^4 \rangle}{\langle \Delta \textbf{x}_i  (\Delta t)^2 \rangle^2} - \frac{3}{5}
\end{equation}
for spatial components $\textbf{x}_i$ of the three dimensional trajectory \cite{Shell2005}. Here the 
averaging is performed over all water molecules and all possible time intervals $\Delta t$.
For simple diffusion, the distribution of step sizes is purely Gaussian, and $\alpha_{\textbf{x}_i}(\Delta t)= 0$ by construction.

When the electric field induces a linear response, it is highly informative to study the time evolution of the transport coefficients using the Green-Kubo approach \cite{Kirkwood}. Accordingly, the viscosity $\eta(t)$ emerges from the time integral of the autocorrelation function $S_{\alpha\beta}=\langle P_{\alpha\beta}(t) P_{\alpha\beta}(0) \rangle$, where $P_{\alpha\beta}$ ($\alpha,\beta=x,y,z$) are the off-diagonal elements of the stress tensor evaluated in simulations:  
\begin{equation}
\frac{k_{\text{B}}T}{V} \eta_{\alpha\beta}(t) =  \int_0^{t} S_{\alpha\beta}(t^\prime) \mathrm{d}t^\prime
= \int_0^{t} \langle P_{\alpha\beta}(t^\prime) P_{\alpha\beta}(0) \rangle\: \mathrm{d}t^\prime\:. \label{eq:GK}
\end{equation}
Here $V$ is the volume of the simulation cell, $k_{\text{B}}$ is the Boltzmann constant, and $T$ is the absolute temperature of the system. Components of the viscosity are obtained by finding the limit
\begin{equation}
\eta_{\alpha\beta} =\lim_{t\rightarrow \infty} \eta_{\alpha\beta}(t)\:.
\end{equation}

Appropriate averaging of components, following symmetry considerations, yields $\eta_\perp$ and $\eta_\parallel$. The apparent shear viscosity $\eta_{\text{app}}$ of water in the presence of the $E$-field is then calculated
\begin{equation}
\frac{1}{\eta_{\text{app}}} = \frac{1}{3} \left( \frac{2}{\eta_\perp} + \frac{1}{\eta_\parallel} \right)\:. \label{eq:eff-eta}
\end{equation}
providing an average viscosity that could be experimentally tested. 

The linear response is obtained for small fields \cite{Sutmann, Svishchev} (SI-Fig.~1b-d), allowing us to calculate $\eta(t)$ for $E=0.0, 0.2$, $0.4$ and $0.6$ V/nm. The time dependent viscosity was extracted from simulations comprising $N_{\text{w}}=23411$ water molecules. Following the equilibration protocol described above, we performed production runs (120 ns at $E=0$ and 300 ns for all $E\neq0$) in the NVT ensemble ($T=300$ K). The update of the neighbour list and the components of the stress tensor were saved to disk at every step. Due to extensive sampling (large system size and long trajectories), the plateau in $\eta_{\alpha\beta}(t)$ is reconstructed with high statistical accuracy up to 50 ps dirrectly from the simulation data,  providing an excellent estimate for the value of the viscosity $\eta_{\alpha\beta}$ in the limit of $t\rightarrow\infty$. The time evolution of $\eta(t)$ was analysed using a self-implemented Kohlraush fit procedure \cite{Kohlrausch_orig,history_Kohlrausch} (see SI section 3 for details). The difference in  $\eta_{\alpha\beta}$  obtained from  the plateau in $ \eta_{\alpha\beta}(t)$ and the extraction from the Kohlraush fit are of the order of $1\%$.

\textbf{\emph{Reorientational dynamics:}} The orientational properties of water molecules were studied using trajectories generated in simulations of static structure. The correlated reorientations of water molecules and their first hydration shells were studied using a self-developed bilateral filtering procedure applied to the time evolution of the projection of the water and shell dipole moment on the field axis (see SI section 4 for details).

\section{\label{sec:Water_Structure}Static structure of water in an electric field}

\begin{figure*}[]
\centering
  \includegraphics[width=\textwidth]{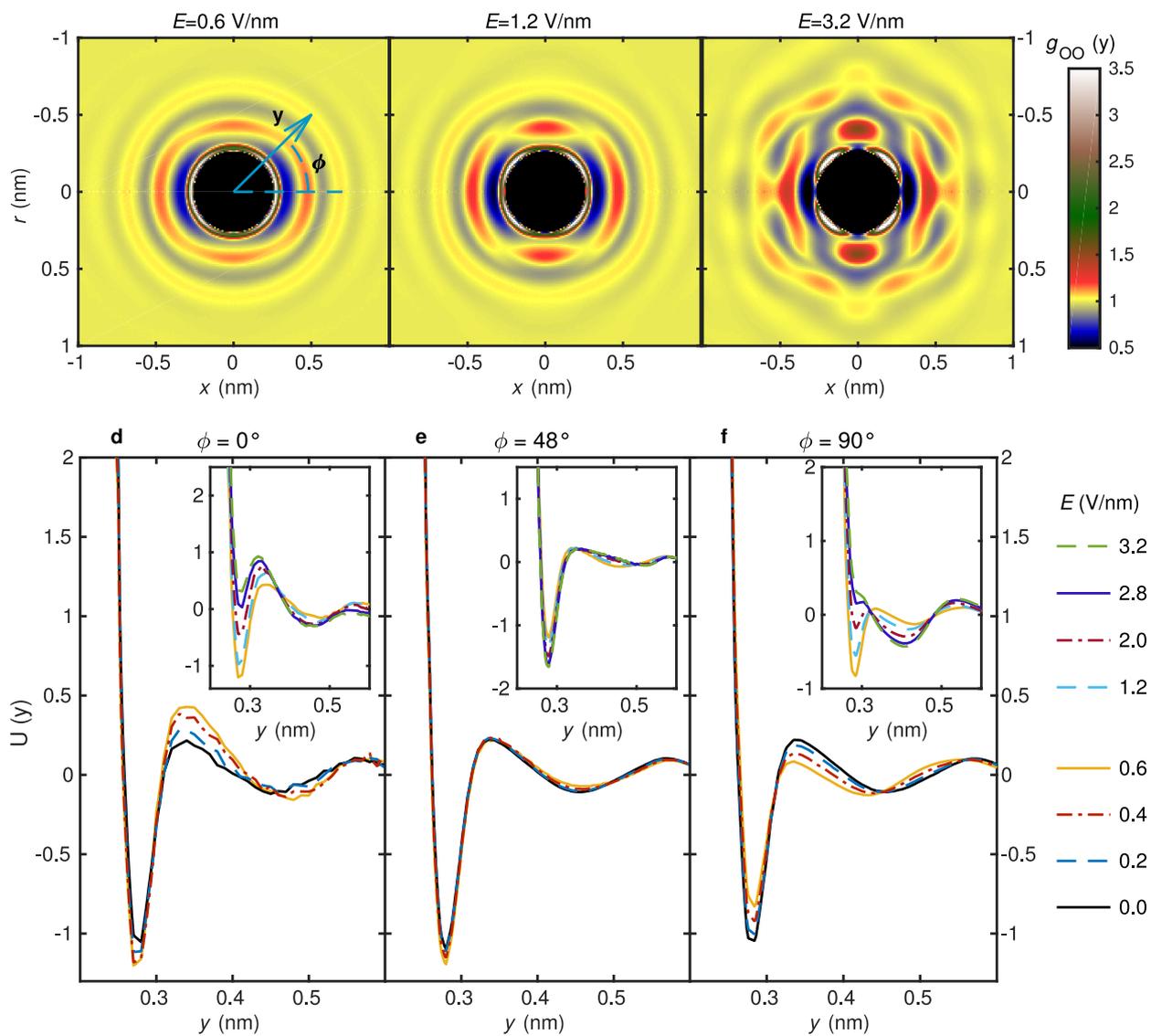}
  \caption{\textbf{Evolution of the static structure of water in the electric field.} Two-dimensional oxygen-oxygen density distribution function $g_{\text{OO}}(r,x)$ for the SPC/E water model at (\textbf{a}) weak fields - $E=0.6$ V/nm, (\textbf{b}) moderate fields - $E=1.2$ V/nm and (\textbf{c}) strong fields - $E=3.2$ V/nm. The distribution is axially symmetric with respect to $r=0$. An arbitrary axis $y$ at the inclination $\phi$ relative to the field acting along the +x axis is noted on the image.  
     Directional potential of mean force $U(y)$ with (\textbf{d}) $\phi=0^\circ$, (\textbf{e}) $\phi=48^\circ$ and (\textbf{f}) $\phi=90^\circ$ are shown for various field strengths. }
  \label{fig:2d_rdf}
\end{figure*}

On the fundamental level, water dipoles exhibit a preference to align with the field in order to optimize the dipole-field interaction energy \citep{Sutmann}, although fields above 3.5 V/nm were found to induce the dissociation of a water molecule \cite{Saitta2012}. This alignment is in competition with the tetrahedral structure of bulk water as well as entropy. Actually, long range order associated with the cuboid lattice, occurs only for very strong fields between 25 and 40 V/nm in computer simulations of classical water  \citep{Sutmann, English2003}. This ordering is perceived as layering of water molecules in both directions, along and perpendicular to the field, where, in the latter case, the layers are rotated by 90 degrees relative to each other. Interestingly, however, no significant restructuring, evidenced by no appreciable change of the radial distribution function of water was found for small fields, and the tetrahedral arrangement of neighbouring molecules  was suggested to be maintained at least until 1 V/nm \cite{Kiselev1996, Zong2016}. Nonetheless, the evolution of the distribution of the angle that water dipoles close with the $E$-field (dipole angle shown in SI-Fig.~2) suggests a gradual transition from a disordered to an ordered phase \cite{Sutmann}, which could be reflected in the emergence of local order at lower field strengths. Nonetheless, reports of the evolution of the water structure under the field are missing.

The static organisation of liquids is typically extracted from the radial distribution function. In agreement with previous reports, we find the mean densities of the first and second hydration shells nearly constant, suggesting that the overall number of neighbours does not change significantly with the field (SI-Fig.~3). However, within the fluid, the imposition of the field induces a symmetry change from isotropic to azimuthal, and the appropriate way to determine changes in water organisation should involve a two-dimensional static density distribution function $g(r,x)$. It is parametrized by $r$ denoting the polar axis ($r \perp E$), and the longitudinal axis $x$ co-aligning with the field. The center of the coordinate system coincides with an atom (O or H) of the central water molecule. Naturally, the azimuthal rotation is out of the plane of the paper. 

The representations shown for $g_{\text{OO}}(r,x)$ in Fig.~\ref{fig:2d_rdf}a-c and respectively for $g_{\text{OH}}$ and $g_{\text{HH}}$ in SI-Fig.~4, permits following the emergence of a significantly more complex organisation of water in the first two hydration shells, while the third shell remains unaffected for low and intermediate field strengths. This anisotropy is perhaps  best captured by observing changes along the spatial angle $\phi$ ( $0^{\circ}<\phi<180^{\circ}$ ), which is the angle between the symmetry axis $x$ and the axis $\mathbf{y}(\phi)$, the latter having its origin at the central atom at $r=x=0$ and a length $y$ (Fig.~\ref{fig:2d_rdf}a and directional density distributions shown in SI-Fig.~5).

\textbf{\emph{Low fields - E $\mathbf{\leq 0.6}$} V/nm:}
Immediately with the imposition of the field (Fig.~\ref{fig:2d_rdf}a), the hydration shells become elliptical, moving outward  in the directions of the field, and inward perpendicular to the field. Two cap-like compartments of increased density appear at $\phi = 0^\circ$ and $180^\circ$, and a depleted ring-like compartment centred at $ 90^\circ$ forms in the first shell. Second hydration shells show exactly opposite trends. The caps and the ring meet at $\phi = 48^\circ$. As the field increases, the caps are filling up until the density inside is nearly constant at around $E= 0.6$ V/nm. This marks the onset of the shell restructuring that is associated with stronger deviations from the linear response (SI-Fig.~1b-d). (For further discussion, see section 1 in the SI.) 
 
These changes are reflected in the free energy landscape for water-water interactions (profiles along particular axes shown in Fig.~\ref{fig:2d_rdf}d-f). While in the cap compartments the first and the second free energy minimum deepen and the barrier for crossing between the shells increases, perpendicular to the field the difference between the first and second shell becomes smaller and smaller. This facilitates anisotropic transport already for small fields.

\textbf{\emph{Moderate fields} - $\mathbf{0.6}$ V/nm $\mathbf{<}$ \emph{E} $\mathbf{ \leq 2.8}$ V/nm:}
At the onset of this regime, the polarization of water starts to saturate, and the electrostriction induces significant deviations from linear response (SI-Fig.~1b-d), fully in agreement with previous work \cite{Sutmann}. In terms of compartmentalization, the most dramatic change, compared to the low-field structure, is the appearance of two new density-enhanced rings in the first hydration shell (centred between $\phi=45^\circ-50^\circ$, and $135^\circ-130^\circ$ (Fig.~\ref{fig:2d_rdf}b). With the field, these compartments densify on the expense of the caps, which now become regions of depletion.  Likewise the density redistribution in the first and second shell stop to coincide in $\phi$ 
(Fig.~\ref{fig:2d_rdf}b), as the water is in part moving from the first into the second shell. Furthermore a series of modifications occur within the compartments as the field increases (see SI section 1 for details).

This redistribution of water is naturally reflected in changes of the free energy distribution around the central water in the field (Fig. 1d-f). The toroidal-shaped minima at $\phi=45^\circ-50^\circ$, and $135^\circ-130^\circ$ are presumably crucial for the maintenance of the tetrahedral ordering reported previously in this regime \cite{Zong2016}. However, a particularly important consequence of the restructuring is the destabilization of the first hydration shell. This must have a major effect on the translation and the rotations of neighbouring waters. Along the axis parallel to the field, remnants of the first shell are visible, but structurally metastable already from $E=2.0$ V/nm on. For these field strengths, the first shell fully merges with the second shell also perpendicular to the field ($\textbf{y}(90^\circ)$).  At this point the molecules have the full freedom to explore the entire 0.55 nm thick ring around the central water.

\textbf{\emph{High fields - E $\mathbf{> 2.8}$} V/nm:}
Further increasing the field strength only enhances the compartmentalization in the first and second shells without new restructuring. However, correlations start to extend significantly deeper into the fluid, as evidenced by the appearance of seven compartments in the third shell (Fig.~\ref{fig:2d_rdf}c). Interestingly, this structuring is highly anisotropic, with the main meridian compartment at $\phi = 90^\circ$ being displaced significantly outward.

Despite the accordance with previous studies  \cite{Sutmann, Zong2016} the high strength of the electric field will cause effects not caught by classical MD simulations, most importantly the spontaneous dissociation of water molecules starting at E = 3.5 V/nm as was shown by AIMD simulations \cite{Saitta2012}.  In the context of experimental confirmation of here-reported results, additional difficulties may arise from electro-vaporization or electro-freezing \cite{English2015,Sutmann,Svishchev,Svishchev1996} at the contact with the electrodes. 

\begin{figure}[]
  \includegraphics[scale=0.78]{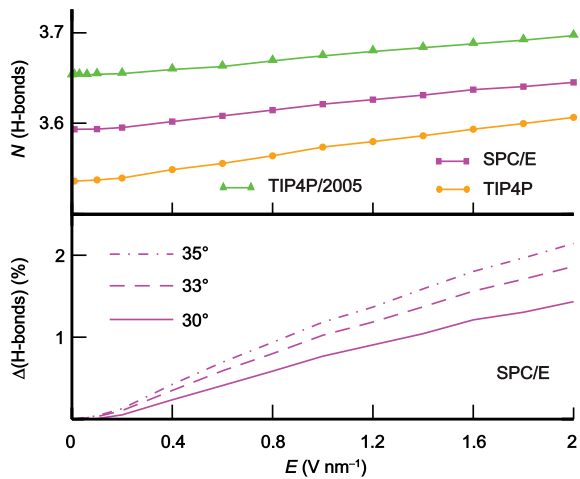}
  \caption{\textbf{Average number of H-bonds per water molecule} as a function of the electric field strength for the three studied water models. The cutoff distance for the H-bond is set to 0.34 nm, while the cutoff angle $\alpha$ = 30$^{\circ}$ (top panel). The change in the number of H-bonds per water molecule for $\alpha$ = 30$^{\circ}$, 33$^{\circ}$ and 35$^{\circ}$ (bottom panel), showing that the contribution of somewhat more distorted H-bonds increases with the field strength.}
  \label{fig:hbonds}
\end{figure}

\textbf{\emph{Hydrogen bonding:}}
Despite the strong compartmentalization, the local tetrahedral structure of water can be maintained (Fig.~\ref{fig:2d_rdf}a-c), especially for low and moderate fields, as was suggested previously \cite{Zong2016}. Our findings suggest that this is enabled by increasing the distortion of (H)-bonds and by the large spatial angle enclosing the compartments within the hydration shells. Actually, the stability and distribution of the H-bonds is affected in an anisotropic manner  as soon as $E>0$ (Fig.~\ref{fig:hbonds}). The probability for establishing H-bonds parallel to $E$ increases with the field strength, and the respective probability for the formation of H-bonds perpendicular to the field decreases \cite{Zong2016}. However, our analysis of the H-bond network shows that the number of H-bonds per water molecule and their angular distortion increases by a couple of percentage points with the increase of the field (Fig.~\ref{fig:hbonds}), suggesting an overall stiffening of the water network. Naturally these effects, occurring on the molecular scale, should be reflected in the time-dependent transport properties of bulk water, such as its diffusion coefficients and viscosity.

\section{\label{sec:Stress_Relaxation}Anisotropy of transport coefficients in the long time-limit}

\begin{figure*}[]
\centering
 \includegraphics[scale=0.85]{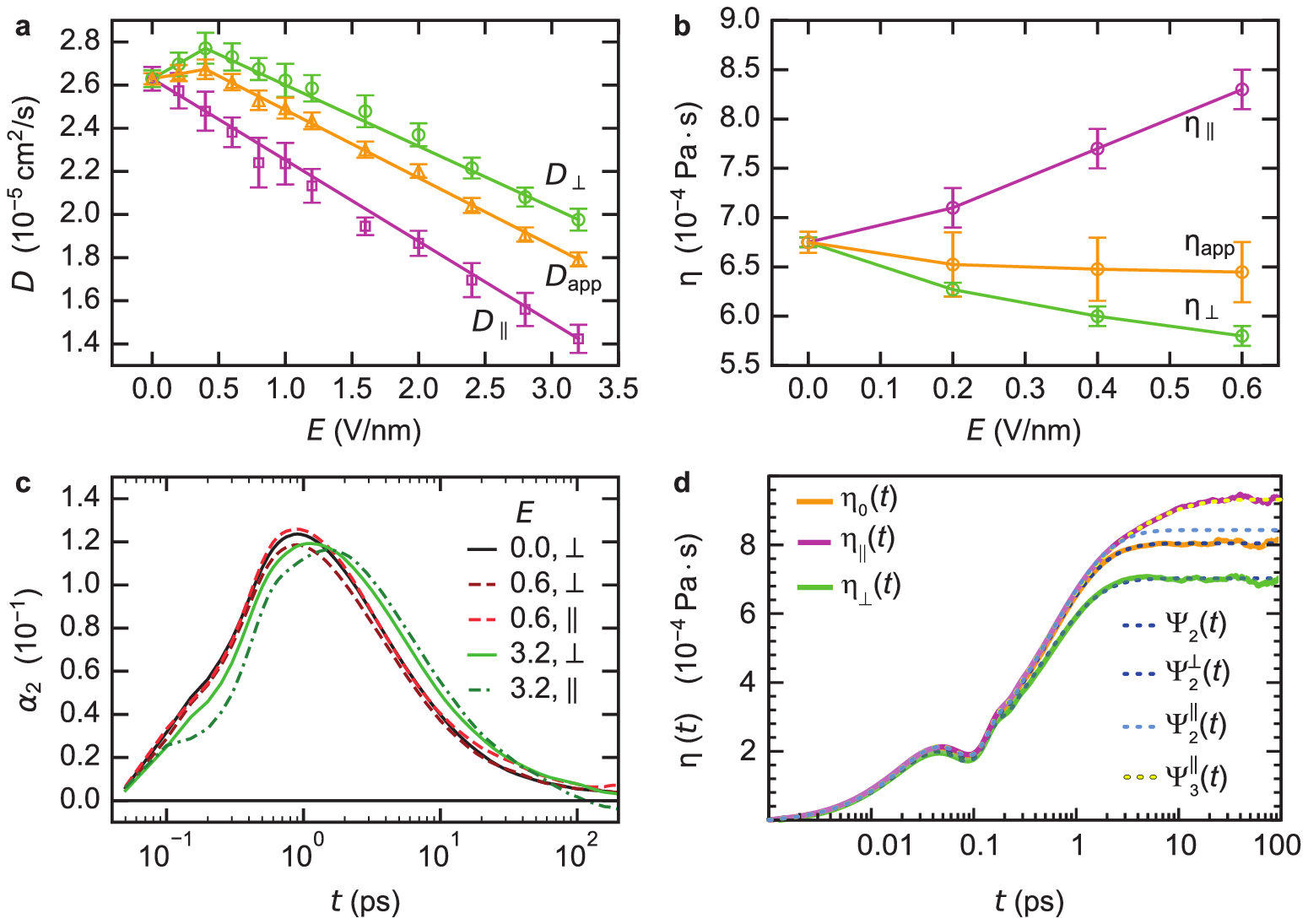}
\begin{tabular}{lccccccc}
& \multicolumn{1}{c}{$E=0$} & \multicolumn{2}{c}{$E=0.2$ V nm$^{-1}$} & \multicolumn{2}{c}{$E=0.4$ V nm$^{-1}$} & \multicolumn{2}{c}{$E=0.6$ V nm$^{-1}$} \\
 Model & \multicolumn{1}{c}{$\eta_0$}  & \multicolumn{1}{c}{$\eta_\perp$} & \multicolumn{1}{c}{$\eta_\parallel$} & \multicolumn{1}{c}{$\eta_\perp$} & \multicolumn{1}{c}{$\eta_\parallel$}  & \multicolumn{1}{c}{$\eta_\perp$} & \multicolumn{1}{c}{$\eta_\parallel$}  \\ 
\hline 
TIP4P       & $4.71\pm0.04$  & $4.54\pm0.05$  & $4.9\pm0.1$  & $4.4\pm0.1$ & $5.1\pm0.2$ &  $4.3\pm0.1$  &  $5.5\pm0.2$  \\           
SPC/E       & $6.75\pm0.05$  & $6.27\pm0.07$  & $7.1\pm0.2$  & $6.0\pm0.1$ & $7.7\pm0.2$ &  $5.8\pm0.1$  &  $8.3\pm0.2$  \\           
TIP4P/2005  & $8.1\pm0.1$    & $7.8\pm0.1$    & $8.5\pm0.2$  & $7.2\pm0.1$ & $9.3\pm0.2$ &  $7.0\pm0.1$  &  $9.3\pm0.2$  \\
\end{tabular}
  \caption{\textbf{Anisotropic transport properties of water.} For the SPC/E model, we show: 
(\textbf{a}) The apparent self-diffusion coefficient ($D_{\text{app}}$) and its components parallel ($D_{\parallel}$) and perpendicular ($D_{\perp}$) to the field as a function of the field strength. 
(\textbf{b}) The apparent shear viscosity $(\eta_{\text{app}})$ and its components in the directions perpendicular $(\eta_{\perp})$ and parallel $(\eta_{\parallel})$ to the $E$-field, in the linear response regime. Both in (a) and (b), the solid lines are guides to the eye.
(\textbf{c})  Time evolution of the non-Gaussian parameter for different field strengths and directions.
(\textbf{d})  Lin-log representation of the time evolution of the isotropic shear viscosity at $E = 0$ (full orange line) and of the shear viscosity parallel (full magenta line) and perpendicular (full green line) to the field at $E = 0.6$ V/nm for the TIP4P/2005 model. The Kohlrausch fits obtained by using $\Psi_2 (t)$ and $\Psi_3 (t)$, are also shown with blue dashed and yellow dashed lines, respectively. The table provides the shear viscosity ($10^{-4}$ Pa$\cdot$s) in the absence of the field $\eta_0$, as well as $(\eta_{\parallel})$ and $(\eta_{\perp})$ in the presence of low $E$-fields. The experimental value at $E=0$ is $8.54\times10^{-4}$ Pa$\cdot$s as stated in NIST Chemistry WebBook. http://webbook.nist.gov/chemistry/fluid (accessed 2018).}
\label{fig:tip4p2005_split}
\end{figure*}

The anisotropy of self-diffusion coefficients \cite{Evans1987, Sutmann, Kiselev1996} and the related viscosity \cite{Zong2016} has been well established for the SPC/E model in the long time limit.
In agreement with this previous work, we find that the parallel component of the diffusion coefficient monotonously decreases with the field (Fig.~\ref{fig:tip4p2005_split}a for SPC/E and SI-Fig.~6 for TIP4P and TIP4P/2005). At the same time, while within statistical uncertainty and with some variability over different water models, the perpendicular component seems to increase for low field strengths, a trend which is reversed at field strengths larger than $E=0.6$ V/nm (Fig.~\ref{fig:tip4p2005_split}a and SI-Fig.~6). We can associate this trend change with the structural reorganisation of the first hydration shell, when the density maxima split and move from $\phi=0^{\circ}$ and $\phi=180^{\circ}$ to $\phi=45^{\circ}$ and $\phi=135^{\circ}$. 

To certify this behaviour at low fields, we make use of the the linear response formalism \cite{Kirkwood} and calculate the anisotropic shear viscosity. The approach is validated at $E=0$, where the obtained viscosity $\eta_0$ (first column in the table provided in Fig.~\ref{fig:tip4p2005_split}) is in full agreement with previous reports \cite{Prosmiti,Wensink,Hess,Chen,Voth}. Accordingly, the TIP4P/2005 model most accurately predicts the experimentally measured $\eta_0=8.54\times10^{-4}$ Pa$\cdot$s at $T=300$ K. In the presence of the field, all water models predict a small decrease in $\eta_\text{app}$ (total change of $2-6\%$), which is a trend that unambiguously changes already for moderate fields, confirming the non-monotonous behaviour of transport coefficients perpendicular to the field and the anisotropy between the two directions, which ranges from $22\%$ to $30\%$ at the critical field strength of $E=0.6$ V/nm (Fig.~\ref{fig:tip4p2005_split}b).

While this long time behaviour clearly shows that the structure affects the dynamics of water in $E$-fields, the understanding of this process emerges from analysing shorter time scales. For example, the long term diffusion coefficients emerge from the analysis of the water mean  square displacements, which are linear already on very short time scales. However, the associated non-Gaussian parameter (Fig.~\ref{fig:tip4p2005_split}c) shows clear deviations from simple diffusion on time scales of up to 10 ps implying much more complex processes governing transport. Interestingly, the non-Gaussian parameter (Fig.~\ref{fig:tip4p2005_split}c) shows anisotropic behaviour already for small field strengths and an appearance of a shoulder on the time scale of $0.2-0.4$ ps at high field strengths in the direction parallel to the field. While these features could be probed experimentally, further efforts are necessary to associate them with particular molecular relaxations.

\section{Characteristic relaxations of the time-dependent transport coefficients}

As discussed in the introduction of this manuscript, two stress relaxation processes have been considered in liquid water in the regime of linear response. The first is a fast one associated with the oscillatory stretching of the H-bonds (time scale of 10 fs). The second, slow process is the restructuring of H-bonds occuring on the time scale of 100 fs. The latter process was recently hypothesized to cause the observed anisotropy of transport coefficients \cite{Zong2016}, while earlier works debated the role of the coupling between rotational and translational degrees of freedom \cite{Evans1986}.

For low field strengths, it is established that the characteristics of these two relaxations can be extracted using Kohlrausch fitting of the normalized stress autocorrelation function $S_{\alpha\beta}(t)/S_{\alpha\beta}(0)$ (Fig.~\ref{fig:tip4p2005_split}d) \cite{Guo,Medina2011,Prosmiti}. The fit involves an empirical two-step relaxation function
\begin{align}
\Psi_2(t) = &\: (1-C_s) \exp \left[ -(t/\tau_{f})^{\beta_f} \right] \cos(\omega t) \nonumber \\
        + &\: C_s \exp\left[ -(t/\tau_{s})^{\beta_s} \right]\:,
\end{align}
where $(1-C_s)$ and $C_s$ are the fractions of the oscillatory fast relaxation and the slow restructuring process described by the Kohlraush decays \cite{Kohlrausch_orig}. Parameters $\tau_f$ and $\tau_s$ are the respective decay times, $\beta_f$ and $\beta_s$ are the expected  non-exponential coefficients.

In several computational studies at $E=0$, the fast excitation is succesfully associated with the  stretching of the H-bond in the terahertz regime \cite{Heyes,Vallauri,Marti,Guo}. Thereby obtained results were in very good agreement with numerous Raman and infrared spectroscopy measurements, as well as inelastic neutron scattering experiments,  despite the simplictiy of the classical water models.  Accordingly, the parameter $\omega$ is the characteristic frequency of the fast  H-bond stretching providing a resonance frequency of the shear viscosity at around 34 ps$^{-1}$ (corresponding to 180 cm$^{-1}$) at $E=0$ \cite{Kauzmann,Walrafen1964,Walrafen1986,Draegert,Dahlborg}.

The slow H-bond recombination is due to the interplay of diffusion and reaction kinetics not a simple, but a stretched exponential, complicating the interpretations of parameters $\tau_s$ and $\beta_s$. However, the average decay time for each process corresponds to the experimental observations for the processes in question \cite{Guo,Medina2011}.

In our simulations at $E=0$, due to excellent statistical accuracy of our simmulation data, we find that $\Psi_2$ fits the data very well over five orders of magnitude in time. The time dependent viscosity, which emerges upon integration of $\Psi_2$ accurately plateaus to $\eta_0$ (Fig.~\ref{fig:tip4p2005_split}d and SI-Fig.~7), which reproduces earlier results of other groups \cite{Guo,Medina2011,Phillips,Li,Laslett,Staib,Shekhar,Milischuk}. Furthermore, we find values of the wave vectors  for the fast oscillations between 219 and 236 cm$^{-1}$, depending on the water model, which is in agreement with previous computational reports \cite{Heyes,Vallauri,Marti,Guo}.

\begin{figure*}[t]
\includegraphics[scale=1]{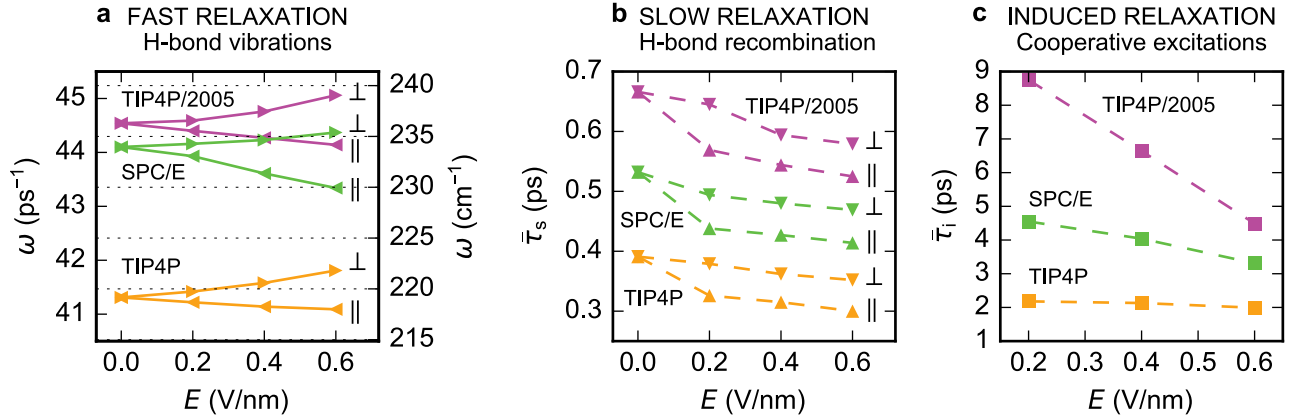}
\caption{\textbf{Effect of small fields on the relaxation processes in liquid water parallel ($\parallel$) and perpendicular ($\perp$) to the $E$-field direction.} 
(\textbf{a}) Changes in the stretching frequency of the H-bonds. (\textbf{b}) Dependence of the average decay times $\bar{\tau}_s$ (eq. (\ref{eq:avg_decay})) on the electric field strength for the slow relaxations of the H-bond network.
(\textbf{c}) Characteristic relaxation time $\bar{\tau}_i$  of the field-induced excitations. Graphs are complemented by the data in SI-Table~5.}
\label{fig:avg_decay}
\end{figure*}

Similarly good correspondence between the fit and the simulations is obtained in the presence of an electric field for the perpendicular component of the time dependent viscosity. The component parallel to the field direction is, however, accurately reproduced only up to several picoseconds, when $\Psi_2 (t)$ starts to plateau prior to the onset of saturation in the simulation data (Fig.~\ref{fig:tip4p2005_split}d). Indeed, for all water models and all field strengths, $\eta_\perp (t)$ and $\eta_0 (t)$ converge on very similar time scales, while the correlations in $\eta_\parallel (t)$ persist significantly longer (Fig.~\ref{fig:tip4p2005_split}d). This hints to the existence of an additional relaxation process that is field-induced and acts only in the parallel direction with a characteristic time scale larger than 1 ps for low field strengths. 

To quantify this field-induced relaxation, an additional Kohlraush decay with a characteristic amplitude $C_i$, time constant $\tau_{i}$, and non-exponential coefficient $\beta_i$ is introduced, and a three-step fit function
\begin{align}
\Psi_3(t) = &\: (1-C_s-C_i) \exp \left[ -(t/\tau_{f})^{\beta_f} \right] \cos(\omega t) \nonumber \\
        + &\: C_s \exp\left[ -(t/\tau_{s})^{\beta_s} \right]\: + C_i \exp\left[ -(t/\tau_{i})^{\beta_i} \right]\:,
\end{align}
is applied to the appropriate data. Remarkably, the obtained fit fully recovers the correlation function and the viscosity parallel to the field, including the long time limit. This corroborates the existence  of a new, so far not characterized coupling of the field and the H-bond network. Notably, since both eqs. (5) and (6) have a large number of fitting parameters, we performed a stability analysis (see SI section 3). As a result, we find that the statistically accurate dependence of several parameters ($\tau_f$, $\beta_f$, $\beta_s$, $\beta_i$) on the field strength could not be determined and these parameters were therefore fixed. Consequently, the number of fit parameters was decreased to three for eq. (5), and five for eq. (6). Importantly, while the absolute values of the fit parameters vary somewhat among the water models, the general behaviour and the determined time scales are always preserved (SI-Tables~1 - 4 and SI-Figs.~8 - 9). This analysis allows us to follow changes in various relaxations in the regime of small fields.

\textbf{\emph{Fast relaxations}} - The most rapid processes are characterized by the first term in $\Psi_2$ and  $\Psi_3$. With increasing field strength, blue and red shifts are observed in $\omega^{\perp}$ and $\omega^{\parallel}$ (Fig.~\ref{fig:avg_decay}a and SI-Tables~1 - 3), respectively. This is consistent with the softening of the H-bond network perpendicular to the field and its stiffening parallel to the field. Given that this finding is universal for all water models, such splitting of the resonant band could be measurable by neutron scattering or Raman spectroscopy. Fit results also show that the $E$-field does not affect $\tau_f$ and $\beta_f$ in a statistically significant manner and consistently with the thermal nature of excitations of H-bonds, the deviations from exponential decay remain small  ($\beta_f=0.848$, TIP4P/2005). As a result, the time integral $T_f$ over the first term in $\Psi_2$ and $\Psi_3$ (see SI section 3, SI-Fig.~9 and SI-Table~4 for details) does not depend on the field intensity or direction $(T_f^{0}\simeq T_f^{\perp}\simeq T_f^{\parallel})$. This yields a conclusion that fast relaxations do not contribute to the anisotropy of transport coefficients.

\textbf{\emph{Slow relaxations}} - The presence of the field, on the other hand, affects the  restructuring of the H-bond network, which is captured by changes in the stretched exponential decay (second term in  $\Psi_2$ and  $\Psi_3$). As a matter of fact, due to the interplay of diffusion \citep{LuzarNat,Luzar2000} and bond kinetics, the deviations from a pure exponential are particularly significant for this process (at $E=0$, $\beta_s=0.595$, TIP4P/2005). As the behavior of stretched exponentials depends strongly on $\beta$, the comparison of characteristic decays should be done using the average decay time $\bar{\tau}_s (\tau_s,\beta_s)$ \cite{Berberan2005}
\begin{equation}\label{eq:avg_decay}
\bar{\tau}_s(\tau_s, \beta_s) = \tau_s \: \frac{\Gamma(2/\beta_s)}{\Gamma(1/\beta_s)}\;.
\end{equation}
Here $\Gamma(.)$ is the gamma function.

The average decay time shows a systematic decrease with increasing field strength (Fig.~\ref{fig:avg_decay}b and SI-Table~5). Interestingly, this suggests faster restructuring overall, with the effect being twice as strong in the parallel direction compared to the direction perpendicular to the applied field. The difference is most likely due to the force acting on dipoles of water molecules building the H-bond. Naturally, these results imply that the contribution of the H-bond restructuring to the shear viscosity, obtained by integrating the second term in $\Psi_2$ and $\Psi_3$ over the whole time domain, decreases in both directions as a function of weak field strength, with $T_s^{0}>T_s^{\perp}>T_s^{\parallel}$ for $E<0.6$ V/nm (SI-Table~4).

\textbf{\emph{Induced relaxations}} - The newly identified characteristic relaxation is, at low fields, an order of magnitude slower than the H-bond restructuring. Furthermore, $\bar{\tau}_i (\tau_i,\beta_i)$, linearly decreases with the field strength, more significantly for TIP4P/2005  than for TIP4P (Fig.~\ref{fig:avg_decay}c and SI-Table~5).  Interestingly, the integrated contribution of this process  $T_i^{\parallel}$ to the shear viscosity increases with the field strength $(T_i^{0}=T_i^{\perp}=0)$, showing that this new relaxation is vital for the overall behaviour of transport coefficients, and largely responsible for their anisotropy.  Since there is a linear relationship between the $E$-field and the torque on a molecule, these results could implicate coupling between molecular rotations and the field. Furthermore, the picosecond time scale suggests that this relaxation may involve more than one water molecule at a time.  However, to elucidate the origin of this process on the molecular scale, further analysis as performed in the following section, is necessary.

\section{Collective excitations}

Guided by our previous results we undertake a systematic analysis of the coupling between the field and the orientation of water molecules characterized by the projection $P$ of the molecular dipole moment vector onto the field direction
\begin{equation}
P = \frac{\vec{d} \cdot \vec{E}}{|\vec{d}\,| \cdot |\vec{E}|}\,.
\end{equation}
Here $\vec{d}$ and $\vec{E}$ are the vectors of the dipole moment and the electric field, respectively.
For $P = 1$, $ \vec{d}\parallel\vec{E}$ are pointing in the same direction and the energy of interaction is minimal. For $P = -1$, $\vec{d}$ and $\vec{E}$ have opposite orientation and the energy is maximal, hence this states are meta stable. Notably, both of these states are torque free.

If existent, collective motions of water molecules respective to the field should involve reorientation of molecules sharing H-bonds, or the stabilisation of the torque free state with $P=-1$ by H-bonds. In both cases, such effects should be captured by the changes in the mean dipole projection $\bar{P}$ of entire hydration shells. We define the orientation of the $i$-th hydration shell as the average orientation of all $N_i$ molecules it contains 
\begin{equation}
\bar{P}_i := \frac{1}{N_i} \sum_{j=1}^{N_i} P_i^{j}\,.
\end{equation}
Each molecule $j$ (excluding the central molecule) is contributing a projection $P_i^{j}$. Importantly, $\bar P_i<0$ suggests that more then one molecule simultaneously adopts the anti-parallel orientation in that hydration shell.

\begin{figure}
\includegraphics[height=0.7\textheight]{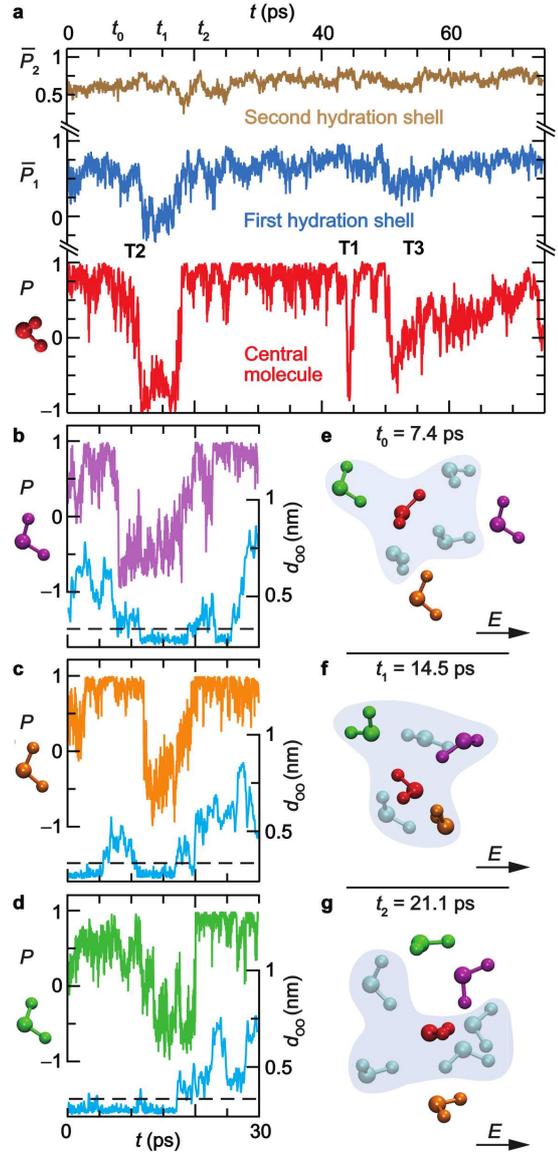}
\caption{\textbf{Cooperative relaxation of the H-bond network.} (\textbf{a}) Time dependent orientation of the central water molecule and the average orientation of waters in its $1^\text{st}$ and $2^\text{nd}$ hydration shell, are shown in red, blue and brown, respectively. The three types of transitions are shown and labelled accordingly. (\textbf{b-d}) Water molecules purple, orange and green, simultaneously adopting an anti-parallel orientation to the field between $t = 0$ and $t = 30\:\text{ps}$. The distance $d_{oo}$ between the oxygen of the central red molecule and the oxygen of the molecule of choice is shown in blue. The horizontal dashed lines mark the end of the first hydration shell ($d_{\text{OO}}<0.34$ nm). Simulation snapshots of the water molecules before reorientation (\textbf{e}), upon coherent motion (\textbf{f}) and back in the parallel orientation (\textbf{g}). Other water molecules that are participating in the first hydration shell (light-blue shaded area in each frame) are shown in blue.}
\label{fig:cooperative_relaxation}
\end{figure}

The time evolution of $P(t)$ for a single water molecule (red trace in Fig.~\ref{fig:cooperative_relaxation}a) shows, that water is predominantly oriented in the direction of the field ($P>0$). However, an occasional stochastic excitation rotates the molecule in the direction opposite to the field. The re-orientation of a given central water may be reflected in the mean orientation of its entire first ($\bar P_1$), but not its second ($\bar P_2$), hydration shell (blue and brown traces, respectively  in Fig.~\ref{fig:cooperative_relaxation}a).  This suggests that simultaneous transitions take place only between nearest neighbours.

To extract lifetimes of excitations from $P(t)$ traces, we used a bilateral filter (see SI section 4 for details). The distribution of these lifetimes is fitted with stretched exponentials
\begin{equation}
p_n(t) = p_n^0 \cdot \exp[-(t/\tau_n)^{\beta_n}]\,,
\end{equation}
allowing for the calculation of characteristic life times $\bar\tau_{n}$ of the transitions using eq.~(\ref{eq:avg_decay}). The index $n$ here denotes a type of a transition since, based on the behaviour of P and $\bar P_1$, several families of excitations  could be characterised, as discussed below.

\begin{figure}
\includegraphics[width=0.4\textwidth]{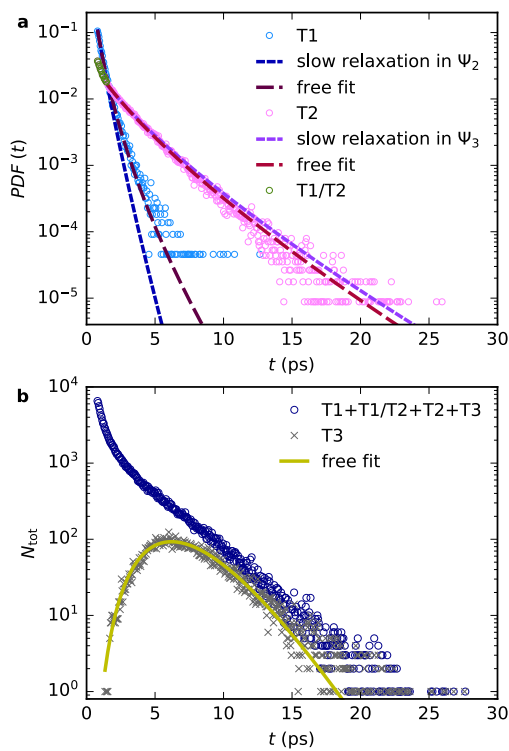}
\caption{\textbf{Distribution of transition times for the TIP4P water model at $E=0.6$ V/nm.} (\textbf{a}) Analysis of T1 and T2 reorientation transitions. Histograms associated with the probability distribution functions (PDFs) of T1, T2 and mixed T1/T2 transition lifetimes, as extracted from simulations, are shown with turquoise, magenta and green symbols, respectively. Stretched exponentials obtained for the slow and the induced process in the stress autocorrelation function are shown with dotted blue and purple lines, respectively. Free fits to the T1 and T2 lifetime probability distributions are shown with dashed brown and red lines, respectively. (\textbf{b}) Analysis of T3 reorientation transitions. A total of 148275 transitions have been identified to construct the probability distribution of all transitions (blue circles). A histogram yielding the T3 life time distribution, as extracted from simulations is shown with gray diagonal crosses, while the free fit by the gamma distribution is shown with the yellow full line.}
\label{fig:lifetime_distribution}
\end{figure}

\textbf{\emph{Transitions of Type 1}} - \textbf{T1} - are identified as spikes in $P$ reaching values smaller than zero, with no signature in $\bar P_1$, as exemplified in Fig.~\ref{fig:cooperative_relaxation}a, where a 3 ps long transition can be seen in $P$. Moreover, a detailed analysis of $P_1^{j}$ traces shows that none of the water molecules in the first hydration shell exhibit a transition at the same time as the central molecule. Relative to other transition types described below, T1s are relatively fast. Hence, very little molecular exchange between the first  and outer shells is observed during the transition, further suggesting that the central molecule establishes H-bonds with its neighbours in the exited antiparallel state, as well as in the relaxed state. Actually, this excitation naturally occurs also in field free water where it is related to the rotations of the molecule within its hydration shell. The reorientation can take place when the H-bonds restructure. Imposing a static electric field, the process becomes faster in the direction parallel to the electric field due to an additional restoring force acting along this particular axis. 

\begin{figure}
\includegraphics[width=0.45\textwidth]{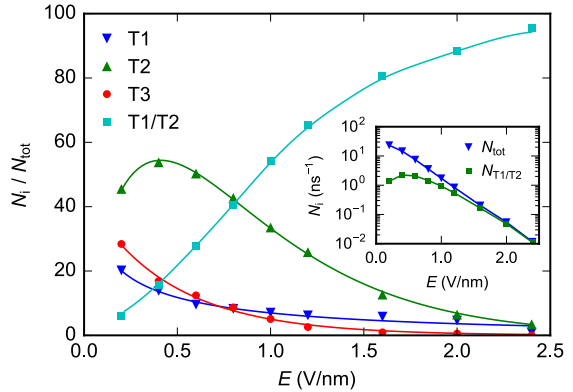}
\caption{\textbf{Relative frequency of transitions as a function of the field strength.} Number of transitions relative to the total number of observed transitions. The inset is showing the total number of transitions and the number of mixed transitions per molecule per nanosecond. The solid lines are guides to the eye. All data is for the SPC/E water model.}
\label{fig:nr_transitions}
\end{figure}

The nature of this transition type, therefore, suggests that the characteristic life time of T1 should compare  to the characteristic time for the slow restructuring of the H-bond network. Actually, we use parameters of the fit of the slow process in the parallel direction and overlay it with the free fit of the distribution of T1 life times (Fig.~\ref{fig:lifetime_distribution}). Remarkably, a very good agreement is obtained between the average decay times obtained from stress relaxations (column 2 in Table \ref{tab:dec_times_comparison}) and those obtained directly from the distributions of transition times (column 3 in Table \ref{tab:dec_times_comparison}) for the entire range of  low field strengths and all water models explored (SI-Table~6). 
This is actually a non-trivial fact as the pressure-pressure correlation function (the integral of which is the average relaxation time) and the distribution of the excitation life times (the integral of which is the average life time) have the same form (or value for the characteristic time). However, the fact that they do, over all field strengths and all water models, show that the stress relaxation associated with the slow relaxation is propagated by restructuring of H-bonds (T1 transitions), as was already established by other methods \cite{Guo,Medina2011,LuzarNat,Marti1996}. Some discrepancies are present, presumably because the analysis of T1 transitions does not recognize the orientational preference of the H-bond restructuring and it is not possible to delineate the random thermal reorientations from the reorientation of molecules to the field. Interestingly, the $\bar\tau_{T1}$ becomes significantly shorter (above $E=1.2$ V/nm), which coincides with the onset of the destabilisation of the first hydration shell (Fig.~\ref{fig:2d_rdf}). While the T1 transitions dominate water dynamics in the absence of the field, under the field they become less and less important, until they basically vanish at already moderate fields (Fig.~\ref{fig:nr_transitions}).

\begin{table}
\begin{tabular}{cccccc}
& \multicolumn{2}{c}{Slow }&\multicolumn{1}{c} {Mixed} & \multicolumn{2}{c}{Induced } \\
$E$  [V/nm] & $\bar\tau_s^{\parallel}$ &$\bar\tau_\text{T1}$ &$\bar\tau_\text{T1/T2}$ & $\bar\tau_\text{T2}$ & $\bar\tau_i$ \\
\midrule
0.2 & 0.44  & 0.55   &0.88     & 6.94    &4.6 \\
0.4 & 0.43  & 0.32   & 0.69    & 3.51    &4.0 \\
0.6 & 0.41  & 0.24   & 0.60    & 2.06    &3.3\\
0.8 &   	     & 0.32   & 0.34    & 1.16    & \\
1.0 &   	     & 0.27   & 0.46    & 0.33    &\\
1.2 &           & 0.24   & 0.40    & 0.13    &\\
1.6 &           & 0.05   & 0.36    & 0.25    &\\
2.0 &           &            & 0.35    &             &\\
\end{tabular}
\caption{\textbf{Average decay times $\bar{\tau}$} in ps (see eq. (\ref{eq:avg_decay})) of the stretched exponential decay. Values are calculated from the fit parameters of the pressure autocorrelation function ($\bar\tau_s^{\parallel}$ and $\bar\tau_i$) and the transition lifetime histogram ($\bar\tau_\text{T1}$, $\bar\tau_\text{T2}$ and $\bar\tau_\text{T1/T2}$). For mixed transition, the entries shifted left and right are indicative of the predominance of T1-like and T2-like transitions, respectively. All data is for the SPC/E water model, TIP4P and TIP4P/2005 water models are shown in SI-Table~6.}
\label{tab:dec_times_comparison}
\end{table}

\textbf{\emph{Transitions of Type 2}} - \textbf{T2} - are identified as sharp excitations in $P(t)$ with $P$ adopting values smaller than zero and a simultaneous transition in $\bar{P}_1$ (Fig.~\ref{fig:cooperative_relaxation}a). The latter is identified by a variant of the watershed algorithm, requiring that the depth of the transition adopts, at least once, a value smaller than $\langle\bar{P}_1\rangle-2\sigma$. Here, the brackets denote averaging of  $\bar{P}_1$ performed over the whole trajectory, whereas $\sigma^2$ is the corresponding variance. This criterion imposes that during a T2 transition $\bar{P}_1$ takes values close to or even below zero (for example at $E=0.6$ V/nm this threshold is 0.367 while the average is 0.676), suggesting a simultaneous reorientation of more than one neighbouring molecule.

T2 transitions clearly have a cooperative character as exemplified in Fig.~\ref{fig:cooperative_relaxation}b-g. The analysis of $P_1^{j}$ traces (Fig.~\ref{fig:cooperative_relaxation}b-d) of molecules in the first hydration shell ($d_{\text{OO}}<0.34$ nm, with $d_{\text{OO}}$ being the distance between water oxygens) shows that they too simultaneously flip into the direction opposite to the field (see $d_{\text{OO}}(t)$ traces in Fig.~\ref{fig:cooperative_relaxation}b-d). This scenario is confirmed  in the  snapshots of the first hydration shell (Fig.~\ref{fig:cooperative_relaxation}e-g) of the red central molecule and the relevant $P_1^{j}$ used in Fig.~\ref{fig:cooperative_relaxation}b-d. Specifically, at $t_0=7.4$ ps all water molecules are pointing in the direction of the field. The central molecule reorients dragging its H-bonded neighbours with it such that at $t_1=14.5$ ps, they all predominantly become aligned opposite to the field, despite the exchange of waters between the first and the second hydration shells (e.g. purple molecule). Finally, by $t_2=21.1$ ps, the entire shell has reoriented and is again parallel with the field. Notably, despite the collective (re)orientations during this entire process, the tetrahedral structure of the hydration shell is, for the most part, preserved in this example. 

A more detailed analysis of T2 transitions shows that at moderate fields, 3 to 6 molecules in average participate in a T2 transition (SI-Fig.~10), but examples with more than 10 participating molecules can be found. The life time of the transition seems to be proportional to the number of participating molecules. At higher field strengths (SI-Fig.~10), the number of participating molecules decreases only for short transitions. In addition, even though cascades of transitions (flipping of molecules in the second, third  hydration shell) do occur, their probability is very low - the tree of transitions does not exceed more that two molecules. The hint to this effect is the lack of transitions in the time evolution of the average dipole moment of the second hydration shell in Fig.~\ref{fig:cooperative_relaxation}a.

As it involves several H-bonded molecules, the flipped state in T2 is energetically more stable than the same number of molecules flipping independently. Furthermore, the energy barrier between the torque free states is higher as the transition state requires simultaneous restructuring of several H-bonds. Even more so than in T1, the H-bonds stabilise the antiparallel orientation of the cluster, evidenced by the still relatively low exchange of molecules between the first and the second hydration shell. Together, this leads to the overall slower dynamics of T2s  compared to T1s. Furthermore, these transitions require the coupling to the field, since it is the electrostatic forces that cause the relaxation of the whole shell back to the original state. In the absence of the field a restoring force is missing and no relaxation would take place. This is indicative of the association of T2 transitions with the induced process observed in pressure correlations at low field strengths. 

The comparison of characteristic decay times obtained by fitting the histogram of T2 life times, and the ones obtained from stress relaxations (Fig.~\ref{fig:lifetime_distribution}), indeed, shows a very good agreement for all field strengths and water models available (columns 5 and 6 in Table \ref{tab:dec_times_comparison} and SI-Table~6). Similarly as for T1, this correspondence is not obvious. However, the fact that it is persistent, over all field strengths and all water models, unambiguously confirms that the newly observed field-induced relaxation of water, responsible for the anisotropy of transport coefficients at low fields, is indeed associated with the cooperative reorientations of water molecules. 

\begin{figure*}
\includegraphics[width=0.95\textwidth]{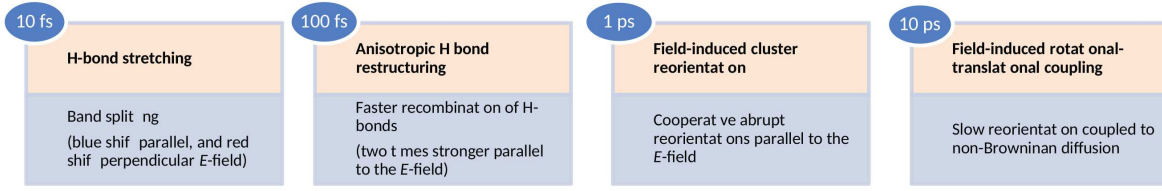}
\caption{\textbf{Summary of processes that couple to the field on respective time scales}. }
\label{fig:overview}
\end{figure*}
Interestingly the fraction of these transitions, which commonly involve several water molecules (such as in Fig.~\ref{fig:cooperative_relaxation}) increases for low fields  (Fig.~\ref{fig:nr_transitions}) until there is a structural change of the first hydration shell (Fig.~\ref{fig:2d_rdf}a). Consequently, at moderate field strengths (see Fig.~\ref{fig:2d_rdf}b), these excitations, involving clusters of molecules, become less and less probable, such that they are no longer detected at high field strengths (Table~\ref{tab:dec_times_comparison} and Fig. 7). 

\textbf{\emph{Mixed transitions}} - \textbf{T1/T2} - Using the criteria associated with T2 transitions, we identify a subset of transitions which show particular characteristics. For low field strengths, these transitions occur on very short time scales and follow the same decay function as the T1 transitions (green symbols in Fig.~\ref{fig:lifetime_distribution}a). More detailed analysis shows that a dominant fraction of these transitions involve flipping of an individual molecule in the presence of a molecule persisting in the antiparallel state in its hydration shell. As such, these are, in essence, T1 transitions. This is confirmed by comparison of characteristic decay times $\bar\tau_{T1/T2}$ and $\bar\tau_{T1}$ for low to moderate field strengths.

Mixed T1/T2s, however, also encompass a subpopulation of  transitions in which two initially bonded molecules simultaneously flip into the antiparallel state, form a transient H-bond, and flip back. Given their cooperative nature, these transitions can be seen as a subclass of T2 transitions.  While being a smaller yet appreciable sub-population in the set of mixed transitions at small field strengths, this subtype becomes dominant at moderate field strengths. At this point the characteristic time $\bar\tau_{T1/T2}$ apparently increases (Table~\ref{tab:dec_times_comparison}). However, this is primarily a consequence of the vanishing  T1-like subpopulation, and the dominance of the T2-like subpopulation.  This is already evident at field strengths of $E=0.6$ V/nm when the average number of molecules participating in a T1/T2 transition is strictly smaller than three but larger than two (see SI-Fig.~10).

Actually, the T2-like events are the only relevant transitions remaining at high fields (Fig.~\ref{fig:nr_transitions}), and especially when the first and the second hydration shells merge (Fig.~\ref{fig:2d_rdf}). However, at this stage the overall number of transitions per nanosecond per molecule drops down by nearly four orders of magnitude compared to the low field conditions (inset in Fig.~\ref{fig:nr_transitions}). 

\textbf{\emph{Transitions of Type 3}} - \textbf{T3} - are identified as cascading signals in $P(t)$ with $P$ adopting values smaller than zero, accompanied by a simultaneous transition in $\bar{P}_1$ (Fig.~\ref{fig:cooperative_relaxation}a), as defined for T2s. To be classified as a T3 transition not more than one state (anti-parallel or aligned) can be reached by an abrupt flip (as shown in the representative example). T3s follow a gamma distribution with an average relaxation time of the order of 10 ps, decreasing with the field strength (Fig.~\ref{fig:lifetime_distribution}b and SI-Table~7). The probability distribution function (PDF) of a gamma distribution is given by
\begin{equation}\label{eq:gamma_dist}
\operatorname{PDF}(t) = \frac{t^{k-1}}{\theta^k \cdot \Gamma (k)}\exp\left(-t/\theta\right)\;,
\end{equation}
where $k$ is a dimensionless shape parameter, $\theta$ is the characteristic relaxation time and $\mu = k\theta$ the average relaxation time.

In a number of cases, T3 transitions are coupled to proximate T2 transitions and are a consequence of the interplay between translational motions and field-coupled rotations. An example of such behavior is shown in Fig.~\ref{fig:cooperative_relaxation}c where, at time $t=t_1$, the orange molecule (exhibiting a T3 transition) is H-bonded to the central red molecule (undergoing a T2 transition). Between $t_1$ and $t_2$, the H-bond breaks and the orange molecule diffuses away into the outer shells of the red molecule. The relaxation of its orientation  parallel to the field may take an extended amount of time because of interactions with the molecules in its new immediate neighbourhood, giving rise to a  $\bar\tau_{T3}$ of the order of 10 ps.  Indeed, this type of coupling between rotational and translational degrees of freedom has been reported in hyper-Rayleigh scattering experiments to occur on similar time scales  \cite{Shelton2000,Shelton2002}. Interestingly, with moderate fields $\bar\tau_{T3}$ increases, while for strong fields, such transitions are prohibited (Fig.~\ref{fig:nr_transitions}).

\section{\label{sec:Conclusion}Discussion and Conclusions}

The coupling between rotational and translational degrees of freedom demonstrated above is a likely source of deviations from simple translational diffusion (Fig.~\ref{fig:tip4p2005_split}c). Remarkably, the non-Gaussian parameter has a non-vanishing behavior on time scales identical to those associated with T3 transitions, suggesting that the analysis of molecular motions in the picosecond regime could yield deeper understanding of water structure. Although existent in the absence of the field, the interplay between the two types of motions can be clearly observed by coupling to a field, as previously suggested \cite{Evans1986}. However, in the context of transport coefficients and their anisotropy, T3 transitions do not appear to have a significant contribution. Namely, T1 and T2 transitions appear to be sufficient for the understanding of the time evolution of viscosity (Fig.~\ref{fig:tip4p2005_split}d). 

Overall, the above presented analysis successfully establishes the relation between the structural organisation of water hydration shells, stress relaxation processes and the water network reorganisation on various time scales. Naturally, the evolution of these processes as the phase transition to a crystalline structure is approached at fields that are even an order of magnitude higher than those explored herein would be informative. However, this would involve circumventing challenges associated with low statistics and slow dynamics, as the number of transitions per water molecule decays rapidly and nonlinear effects are very large.

Using molecular dynamics simulations we simulate an unconstrained block of liquid water in a uniform electric field, a methodology that has proven instrumental to elucidating the physical foundations for a number of the anomalous properties of water. We analysed selected bulk properties over a large range of static electric fields. Our work builds and comments on the simulation literature where the current setup is standard and widely used \cite{Evans1987, Evans1984, Vegiri2004, Evans1986, Vegiri2001a, Vegiri2001b, Vegiri2002, Vegiri2003, Sutmann, Zong2016, Kiselev1996,Jung1999}. In this ensemble of works, field strengths of up to 40 V/nm have been studied \cite{Sutmann}, but indeed, we limit ourselves to fields which are below ones inducing the dissociation of water (3.5 V/nm) \cite{Saitta2012}. Despite the simulated geometry, the results do not imply that the experiment must be performed in a condenser. Actually, our results could be regarded as the zero frequency response. Furthermore, such field strengths can be imposed by applying fields with strengths of several kV on tips of 10 to 100 nm radius. Surely, this is experimentally very challenging and  further problems like dielectric saturation  of electrodes \cite{Szklarczyk1989} and heating complicate the interpretation of the experiment \cite{Scovell2000,English2015}. However, the analysis of the idealized conditions, as performed in our manuscript is necessary for understanding this complex response.

For several common water models we found that the electric field promotes changes in local order responsible for the emergence of structural transitions and the appearance of anisotropic transport coefficients.  Specifically, we show that the anisotropic compartmentalization within the first and second hydration shells can be related to the stability and stiffness of the H-bond network, the latter coupling to the field on  time scales spanning four orders in magnitude, as summarized in Fig.~\ref{fig:overview}.

Detailed analysis of each process, associated with the relaxation of H-bond network and our ability to relate molecular transitions to stress relaxations opens a wealth of possibilities for future experimental verification of the predictions of the current modelling. Naturally, different techniques should be used to probe the response on different time scales. For example, on the time scale of 10 fs, high resolution X-ray spectroscopy has been used to determine the coordination of water molecules in the fist hydration shell \cite{Wernet2004}, while Raman spectroscopy and neutron scattering have proven instrumental in measuring the characteristic frequency of H-bond stretching, which is a band that is predicted to split under the field.   

On the time scale of 100 fs, we find restructuring of H-bonds \cite{LuzarNat}, often studied using various types of X-ray \cite{Hura20001,Hura20002} and infrared spectroscopy \citep{Teixeira,Laenen,Nienhuys}.  Under the field, our modelling predicts a general decrease of the H-bond recombination times, twice as pronounced in the  parallel direction as compared to the perpendicular one. As a result, the difference of relaxation times in the two directions of about $10-20\%$ is expected already at small field strengths, which could be experimentally accessible.

Most exciting, however,  would be the confirmation of the existence of the induced cooperative relaxations occurring on the picosecond time scale. Besides being seen in stress relaxations and on the level of small molecular clusters, its signature can be found  in the time-dependent autocorrelation function of the average polarization vector of the first hydration shell (SI-Fig.~11). As demonstrated in previous sections, these excitations are crucial to explain the anisotropy of transport coefficients. While potentially still accessible to infrared spectroscopy, pump-probe experiments could yield direct observation of individual relaxations in a time-resolved manner \cite{Stolow2004}, using hyper-Rayleigh scattering \cite{Shelton2000,Shelton2002}. These techniques could be also used to study the coupling between translational motions and the electric field occurring on time scales of decades of picoseconds.

Last but not least, a similar response to an applied field may be envisaged for other transport properties that depend on the H-bond network, such as the bulk viscosity or thermal conductivity. In that respect, our study represents a model platform to explore the behaviour of fluids in  symmetry-broken environments and when the invoked effects are small but, nevertheless, important.

\textbf{Acknowledgements} ASS and AB acknowledge financial support by the Bavarian Ministry of Economic Affairs and Media, Energy and Technology for the joint projects in the framework of the Helmholtz Institute Erlangen-N\"urnberg for Renewable Energy (IEK-11) of Forschungszentrum J\"ulich. ZM, DMS and ASS were funded by the Croatian structural funds project MIPOMAT and the FAU Erlangen Cluster of Excellence: Engineering of Advanced Materials.

\textbf{The authors have no competing interest.}

\textbf{Author contributions} A.S.S. and D.M.S. conceived and supervised the work. A.B and Z.M. performed the simulations and the analysis. All authors interpreted the data and wrote the manuscript.




\bibliographystyle{elsarticle-num} 
\bibliography{visco_noURL}





\end{document}